\documentstyle[11pt]{article}
\newcommand{\beq}{\begin{equation}}
\newcommand{\eeq}{\end{equation}}
\def\bra#1{\langle #1 |} \def\ket#1{| #1 \rangle}
 
\def\expect#1{\langle #1 \rangle} \def\d{^\dagger}

\def\half{\frac{1}{2}}

\title{From Quantum to Classical: the Quantum State Diffusion model}

\author{
N.~Gisin, \\
{\protect\small\em Group of Applied Physics,}
{\protect\small\em University of Geneva, 1211 Geneva 4, Switzerland}\\
T.A.~Brun\thanks{New address:  Institute for Theoretical Physics,
University of California, Santa Barbara, CA 93106 U.S.A.}, \\
{\protect\small\em Physics Department, Queen Mary and Westfield College,
University of London,}\\
{\protect\small\em London  E1 4NS, England}\\
M.~Rigo \\
{\protect\small\em Mathematics Department, Royal Holloway College,
University of London,}\\
{\protect\small\em Egham, Surrey TW20 0EX, England}
}
\date{\today}

\begin{document}
\maketitle

\section{Introduction}\label{int}
The vagueness of the quantum-classical border in physics makes some
physicists uncertain about the foundation of their science. Others,
by contrast, feel that the accuracy of the verified predictions makes
these foundations especially solid. This longtime state of affairs is
changing in an interesting way. Recent progress in mesoscopic 
physics, and the so-called ``new science of nanotechnology,''
open new domains to be explored by 
theoreticians and experimentalists alike. In this contribution
we will exploit numerical and conceptual tools that were developed
to explore this new domain of mesoscopic physics, and apply them to
systems that are close to the quantum-classical border.

Central to classical mechanics are the concepts of trajectories and
the phase space structures associated with them.
Hence, we would like to illustrate our results with figures showing
these trajectories and structures as they appear in our models. These figures,
and the fact that they can be computed without any supplementary parameters,
constitute the main result of this contribution.

As examples we could use several kinds of oscillators.
The first, and the simplest, is the harmonic oscillator with
its regular orbits; but this is boring!  Next, systems that exhibit
classical chaos, such as the
Kicked Anharmonic OScillator (KAOS)
\cite{SpillerRalph94,Spiller95,GisinRigo95,RigoGisin96}
and the double-well Duffing oscillator \cite{Brun95a,Brun95b}.
In both cases the structure of the strange attractors
emerges out of quantum cloudiness when the quantum-classical border is
approached. Another example worth mentioning is a nonlinear oscillator
exhibiting hysteresis \cite{RigoAlbert96},
the hysteresis curve appearing when the system
becomes more classical.

The feature which characterizes all these systems is that they are open;
they interact significantly with some kind of external environment.
One cannot in general describe an open system
by a single state vector; one requires a density matrix, essentially
giving classical probabilities for the system to be in given quantum
states.  In the limit of Markovian systems, this density matrix
evolves according to a linear master equation, as we shall see below.

For many reasons, we would like to get away from the density operator
description, and these new tools allow us to do so.
These tools require no additional parameters, only a simple idea:
describe open quantum systems by state vectors, as for closed systems.
Unlike closed systems, their evolution is not unitary---in fact, it
is not linear and not deterministic (two general characteristics
of open systems). The evolution is chosen such that when the pure state
defined by the (normalized) state vector is averaged over the randomness,
the usual density matrix description is recovered at all times. The numerical
advantage of this approach stems from replacing an $n\times n$ matrix by an
$n$ component vector. The conceptual advantage lies in the possibility to have
individual states in the theory, as in actual experiments, and to
compute mean values as averages of individual outcomes, again as in
actual experiments.

This approach shares with the decoherence approach the
idea that the environment plays a crucial role in breaking the unitary
(boring) evolution and reducing the state
\cite{JoosZeh85,Zurek91}.
However, here the evolution
provides classical probabilities for quantum amplitudes. 
Our approach also shares
some similarities with the GRW spontaneous localization model
\cite{GRW}, in that the
evolution equations are stochastic, but preserve pure states. But here
there are no {\it ad hoc} parameters.

\section{Quantum State Diffusion and Quantum Jumps}
Markovian open quantum systems are usually
described by a master equation:
\beq
{\dot \rho} = - i [H,\rho] + \sum_m \biggl( L_m \rho L_m\d
  - {1\over2} \{ L_m\d L_m, \rho \} \biggr)
\label{master_eqn}
\eeq
where $\rho$ is the density matrix for the system, $H$ its Hamiltonian,
and the linear operators  $L_m$ describe the effects of the
environment. In an {\it unraveling} of such master equations,
one describes the system in terms of a normalized
pure state $\ket{\psi(t)}$ which follows a stochastic ``trajectory'' in
Hilbert space.  By averaging the pure state projector $\ket\psi\bra\psi$
over all possible trajectories with appropriate weights, one reproduces
the density operator $\rho = M(\ket\psi \bra\psi)$.  This is analogous
classically to replacing the Fokker-Planck equation for probability densities
with a stochastic Langevin equation for single trajectories.

Unfortunately, unlike the case of classical Brownian motion, the unraveling
of the master equation (\ref{master_eqn}) is not unique.  Thus, there is
some ambiguity in how one separates classical and quantum uncertainties,
related to the ambiguity in identifying density matrices with quantum
ensembles.  In this section we consider two well known unravelings of
(\ref{master_eqn}).

In quantum state diffusion (QSD),
the (It\^o) stochastic evolution equation for the normalized state
vector $\ket{\psi(t)}$ reads:
\begin{eqnarray}
\ket{d\psi(t)} &=&  -i H \ket{\psi(t)} dt
  - \half\sum_j (L_j^\dagger L_j
  - 2\expect{L_j^\dagger}_\psi L_j
  + |\expect{L_j}_\psi|^2) \ket{\psi(t)} dt  \nonumber\\
&&   + \sum_j (L_j - \expect{L_j}_\psi) \ket{\psi(t)} d\xi_j
\label{QSDeq}
\end{eqnarray}
where the ``noises'' $d\xi_j$ are complex-valued Wiener processes of zero mean
$M(d\xi_j)=0$ and correlations
$M(d\xi_j d\xi_k)=0$, $M(d\xi_j^* d\xi_k)=\delta_{jk}dt$.
This equation describes a continuous non-differentiable
evolution similar to the familiar
diffusive paths of a classical Brownian particle, but in Hilbert space
instead of real space.  QSD is the only continuous unraveling which
satisfies the same symmetries as the master equation itself
\cite{GisinPercival,GisinRigo95}.

Our second example is the quantum jumps (QJ) unraveling, which
is closely related to photon counting
\cite{Dalibard92,Carmichael93}.
However, it can be
defined for any master equation (\ref{master_eqn}).
The stochastic increment for the wave-function is
\begin{eqnarray}
\ket{d\psi(t)} &=&  -i H \ket{\psi(t)} dt
  - \half\sum_j (L_j\d L_j
  - \expect{L_j\d L_j}_\psi) \ket{\psi(t)} dt  \nonumber\\
&& + \sum_j \biggl(
  \frac{L_j \ket{\psi(t)}}{\sqrt{\expect{L_j\d L_j}_\psi}}
  - \ket{\psi(t)} \biggr) dN_j
\label{QJeq}
\end{eqnarray}
The discrete Poissonian noises $dN_j$ assume the values 0 or 1.
Most of the time $dN_j=0$ and the evolution is continuous
and differentiable. However, whenever
$dN_j=1$ there is a ``jump'' to the state
$L_j \ket{\psi(t)}/\sqrt{\expect{L_j\d L_j}_t}$.
The $dN_j$ processes have mean values
$M_{\ket\psi}(dN_j)=\expect{L_j\d L_j}_\psi dt$ and correlations
$dN_j dt=0$ and $dNj dN_k=\delta_{jk}dN_j$.  This means, essentially,
that jumps occur randomly with an average rate $\expect{L_j\d L_j}$.

It has been shown that a quantum jump description of balanced heterodyne
detection produces an equation identical to the QSD equation (\ref{QSDeq})
in the limit of a strong local oscillator \cite{Wiseman,GarrawayKnight},
and that a QSD description
of photon detection produces jump-like behavior \cite{GisinPercival}.
Also, in the classical
limit (i.e., high photon numbers), the quantum jump equation begins to
exhibit diffusive behavior as well \cite{BGOR}.
Thus, though these unravelings are
distinct, they can behave similarly in certain limits.

\section{Harmonic oscillator}
Let us illustrate these two unravelings for a simple example: the
damped harmonic oscillator at finite temperature. $H=\omega a^\dagger a$,
$L_1=\sqrt{\bar n\gamma}~a^{\dagger}$ and
$L_2=\sqrt{(\bar n+1)\gamma}~a$ where $\bar n$ is the thermal equilibrium
mean photon  number, $\bar n=\expect{a^{\dagger} a}_\rho$,
and $\gamma$ is the inverse relaxation time.
For QSD one can show that any initial states tends to a
coherent state: $\ket{\psi(t)} \rightarrow
\ket{\alpha_t}$, where $a\ket{\alpha_t} = \alpha_t \ket{\alpha_t}$
and $\alpha_t = (\expect{q}_\psi + i\expect{p}_\psi)/\sqrt{2}$.
Furthermore, the evolution
of $\alpha_t$ is governed by a classical equation,
\beq
d\alpha_t = -i\omega\alpha_t dt - {\gamma \over 2} \alpha_t dt
+ \sqrt{\bar n\gamma} d \xi_t
\label{dalpha}
\eeq
Hence, for this example at least, the QSD equation fully describes how the
environment localizes the quantum state down to a minimum Gaussian wavepacket,
and how this wavepacket follows a classical trajectory
\cite{Percival94,Halliwell95,Schack95,Steimle95,Holland96,GisinBaltimore96}.

For quantum jumps a similar analysis can be carried out, and for a system
in a coherent state the evolution of $\alpha_t$ is governed by the equation
\beq
d\alpha_t = -i\omega\alpha_t dt - {\gamma \over 2} \alpha_t dt
+ \sqrt{\bar n\gamma} {\alpha \over \sqrt{|\alpha|^2 + 1}} dW_t,
\label{dalpha_jump}
\eeq
where $dW_t$ is a real, non-Gaussian stochastic differential variable
with mean $M(dW_t) = 0$ and $M(dW_t dW_t) = dt$,
representing the difference between the mean of the jump
processes and their actual values.  In the classical limit
of large $\alpha$, this $dW_t$ approaches a diffusion process, and
equation (\ref{dalpha_jump}) has behavior essentially identical to
(\ref{dalpha}).

Somewhat surprisingly, states tend towards small wavepackets for quantum
jumps as well as QSD, so the same ``classical limit'' seems to exist for
both \cite{BGOR}.

\section{Chaotic Duffing oscillator}

A good example of a nonlinear system is the forced, damped
Duffing oscillator.
This has a classical equation of motion
\begin{equation}
{d^2x\over dt^2} + 2\Gamma{dx\over dt} + x^3 - x = g\cos(t),
\label{duffing_eom}
\end{equation}
and for some choices of $\Gamma$ and $g$ is chaotic
\cite{Gutzwilleretal}.

Because the equation of motion includes explicit time-dependence, the
solutions lie in a three-dimensional phase space ${x,p,t}$.  It is
helpful to consider a discrete surface of section of this system.
Let $(x_0, p_0)$ be the initial point of the forced, damped Duffing
oscillator at time $t_0=0$.  Then we can define a {\it constant phase map}
in the $x$-$p$ plane by the sequence of points $(x_n, p_n) = (x(t_n), p(t_n))$
at times $t_n = 2\pi n$.  Figure 1 illustrates this in the
chaotic regime, where we can clearly see from the surface of section the
fractal structure of the strange attractor.

Quantizing the Duffing oscillator is straightforward using the QSD formalism.
The Hamiltonian operator is
\begin{equation}
H(Q,P,t) =
  P^2/2m + Q^4/4 - Q^2/2 + g\cos(t) Q
  + \sqrt\Gamma (QP + PQ),
\end{equation}
and the damping is represented by a Lindblad operator
\begin{equation}
L = 2\sqrt\Gamma a = \sqrt{2\Gamma}( Q + iP),
\end{equation}
where we have assumed $\hbar = 1$.
The last term in the Hamiltonian is an ansatz, added to give the correct
equations of motion in the classical limit; it is necessary due to the
simplistic model of the dissipative environment.

This system is far from classical.  To
go to the classical limit, we introduce a scaling factor $\beta$,
\begin{equation}
H_\beta(Q,P,t) = P^2/2m + Q^4/4\beta^2 - Q^2/2
  + g\beta\cos(t) Q + \sqrt\Gamma (QP + PQ).
\label{hamiltonian}
\end{equation}
As we increase $\beta$, the scale
of the problem (compared to $\hbar$) increases by $\beta$ in $x$ and $p$,
without altering the dynamics.  Thus,
$\beta \rightarrow \infty$ is the classical limit of this system.
Classical behavior
should emerge from the system in this limit.  This is
supported by the numerical calculations. (See figures 1 and 2.)

The classical unscaled problem is bounded within a small
region of phase space.  (See figure 1.)
Clearly, the quantized version
of this problem (with $\hbar=1$) should be far from the classical limit.
One would expect to observe little trace of the classical fractal
structure.  The expectation values of $Q$ and $P$ should be
dominated by noise.
We solved the forced, damped Duffing oscillator
numerically by integrating the QSD equation for three different
values of the parameter $\beta$.

Examining figure 2a, we see that the results of the
numerical calculation match our expectations very well.  The expectation
values appear randomly distributed; they are dominated almost
completely by the stochastic terms of the equation.

Approaching the classical limit of large $\beta$,
more and more of the classical structure of the attractor appears.
At first the broad outlines of the attractor are formed, then increasing
levels of substructure.  (Figures 2b--2c.)  The full fractal structure
of a strange attractor is only attainable in an unattainable
purely classical limit.  In an actual physical system, the uncertainty
principle provides a lower cutoff to the scale-invariance
of the strange attractor.  

It can easily be shown that the quantum jumps formalism has a ``diffusive
limit'' similar to QSD as one approaches the classical limit \cite{BGOR}.
Thus, exactly the same qualitative behavior is demonstrated in that case
as well.

\section{Conclusions}
Neither QSD nor QJ can be regarded as a fundamental theory, since both
divide the universe arbitrarily into a system and environment; and they
represent two different and interesting unravelings of the
master equation (though only QSD has the same symmetries as the master equation 
\cite{GisinPercival,GisinRigo95}).
This is similar to the ambiguity which arises in
consistent histories, in which it is possible to choose different sets
of histories corresponding to different coarse-grained descriptions
\cite{DH}.
On the other hand, QSD (or QJ) makes testable predictions, and
demonstrates the rise of classical physics in a way which is difficult
or impossible to see with other theories.

Without entering into a sterile debate, let us note that there seems to
be an interesting link between consistent histories and the stochastic
models used in this contribution.  It appears that the solutions of the
stochastic equations provide consistent sets of histories.
Hence, it seems that the two approaches are related in a way similar
to the relationship in classical mechanics between the global view
given by the Maupertuis principle and the local view given by the
Newton equations.  The global view may be more elegant,
but the local one is more useful to compute actual numbers
\cite{DGHP,Brun96}.

Quantum mechanics is nonlocal. Classical mechanics is local. Consequently
classical mechanics can not explain all quantum phenomena.
Conversely, it is cumbersome to use
quantum mechanics to describe classical phenomena. Not only are the
computations more complex, but---and this is the main point---it is
conceptually more difficult: one has to argue that nonlocality, entanglement
and the principle of superposition can be set aside when crossing the
``quantum $\rightarrow$ classical'' border. Clearly, nonlocality, entanglement
and the principle of superposition should become irrelevant in the classical
limit. But why should one argue? Shouldn't it just come out of the equations?
Does it come out of the equations? This contribution is about the last
question. And the answer is: ``it depends on which equation.''

\section*{Acknowledgments}
We would like to thank Lajos Di\'osi, Francesca Mota-Furtado,
P.F.~O'Mahony, Ian Percival and R\"udiger Schack for useful conversations.
This research was funded in part by the UK EPSRC, the EU Human Capital and
Mobility Program, and the Swiss National Science Foundation.

\vfil

Figure 1.  The constant phase surface of section for the classical forced,
damped Duffing oscillator in the chaotic regime, $\Gamma = 0.125$, $g = 0.3$.

\vfil

Figure 2.  The constant phase surface of section for a single QSD
trajectory of the quantum forced, damped Duffing
oscillator in the chaotic regime, $\Gamma = 0.125$, $g = 0.3$,
for three scalings:  a) $\beta = 1$, b) $\beta = 4$ and c) $\beta = 10$.

\vfil


\begin{thebibliography}{99}

\bibitem{SpillerRalph94} T.P. Spiller and J.F. Ralph,
        Phys. Lett. A, {\bf 194}, 235, (1994).

\bibitem{Spiller95} T.P. Spiller, J.F. Ralph, T.D. Clark, R.J. Prance
  and H. Prance, J. Low Temp. Phys., {\bf 101}, 1037 (1995).

\bibitem{GisinRigo95} N. Gisin and M. Rigo, J. Phys. A, 28, 7375-7390, 1995.

\bibitem{RigoGisin96} M. Rigo and N. Gisin,
        Quantum and Semiclass. Optics 8(1), 255, (1996).

\bibitem{Brun95a} T.A. Brun, Phys. Lett. A, {\bf 206}, 167 (1995).

\bibitem{Brun95b} T.A. Brun, J. Phys. A, {\bf 29}, 2077 (1995).

\bibitem{RigoAlbert96} M. Rigo, G. Alber, F. Mota-Furtado and P.F. O'Mahony,
  submitted to Phys. Rev. A.

\bibitem{JoosZeh85} E. Joos and H.D. Zeh, Z. Phys. B, {\bf 59},
  223 (1985).

\bibitem{Zurek91} W.H. Zurek, Physics Today, 36 (October 1991).

\bibitem{GRW} G.C. Ghirardi, A. Rimini and T. Weber, Phys.~Rev.~D, {\bf 34},
  470 (1986).

\bibitem{GisinPercival} N. Gisin and I.C. Percival, J. Phys. A,
  {\bf 25}, 5677 (1992);
  J. Phys. A, {\bf 26}, 2233 (1993);
  J. Phys. A, {\bf 26}, 2245 (1993).

\bibitem{Dalibard92} J. Dalibard, Y. Castin, and K. M\o lmer,
  Phys. Rev. Lett., {\bf 68}, 580 (1992); see also
  J. Opt. Soc. Am., {\bf 10}, 524 (1993).

\bibitem{Carmichael93} H.J. Carmichael, {\sl An Open Systems Approach
  to Quantum Optics, Lecture Notes in Physics m18}, Springer (Berlin 1993).

\bibitem{Wiseman} H.M. Wiseman and G.J. Milburn, Phys.~Rev.~A, {\bf 47},
  642 (1993).

\bibitem{GarrawayKnight} P.L. Knight and B. Garraway, in {\it Quantum 
Dynamics of Simple Systems},
  eds G.L. Oppo et al., The 44th Scottish University Summer School in 
Physics, pp 199-237 (1996).

\bibitem{BGOR} T.A. Brun, N. Gisin, P.F. O'Mahony and M. Rigo,
  quant-ph/9608038, submitted to Phys.~Rev.~Letters.

\bibitem{Percival94} I.C. Percival, J. Phys. A, {\bf 27}, 1003 (1994).

\bibitem{Halliwell95} J.J. Halliwell and A. Zoupas, Phys. Rev. D, {\bf 52},
  7294 (1995).

\bibitem{Schack95} R. Schack, T. Brun and I.C. Percival, J. Phys. A,
  {\bf 28}, 5401 (1995).

\bibitem{Steimle95} T. Steimle, G. Alber and I.C. Percival, J. Phys. A,
  {\bf 28}, L491 (1995).

\bibitem{Holland96} M. Holland, S. Marksteiner, P. Marte and P. Zoller,
  Phys. Rev. Lett., {\bf 76}, 3683 (1996).

\bibitem{GisinBaltimore96}M. Rigo and N. Gisin, Quantum and Semiclassical 
Optics, {\bf 8},  
  255-268 (1996).

\bibitem{Gutzwilleretal} M.C. Gutzwiller, {\sl Chaos in classical and
  quantum mechanics}, Springer, (Berlin, 1990);
  J. Guckenheimer and P. Holmes, {\sl Nonlinear oscillations, dynamical
  systems, and bifurcations of vector fields}, Springer, (Berlin, 1983).

\bibitem{DH} R. Griffiths, J. Stat. Phys., {\bf 36}, 219 (1984);
  R. Omn\`es, Rev. Mod. Phys., {\bf 64}, 339 (1992);
  H.F. Dowker and J.J. Halliwell, Phys. Rev. D, {\bf 46}, 1580 (1992);
  M. Gell-Mann and J.B. Hartle, Phys. Rev. D, {\bf 47}, 3345 (1993).

\bibitem{DGHP} L. Di\'osi, N. Gisin, J.J. Halliwell and I.C. Percival,
  Phys.~Rev.~Lett., {\bf 21}, 203 (1995).

\bibitem{Brun96} T.A. Brun, quant-ph/9606025, submitted to Phys.~Rev.~Letters.

\end{thebibliography}
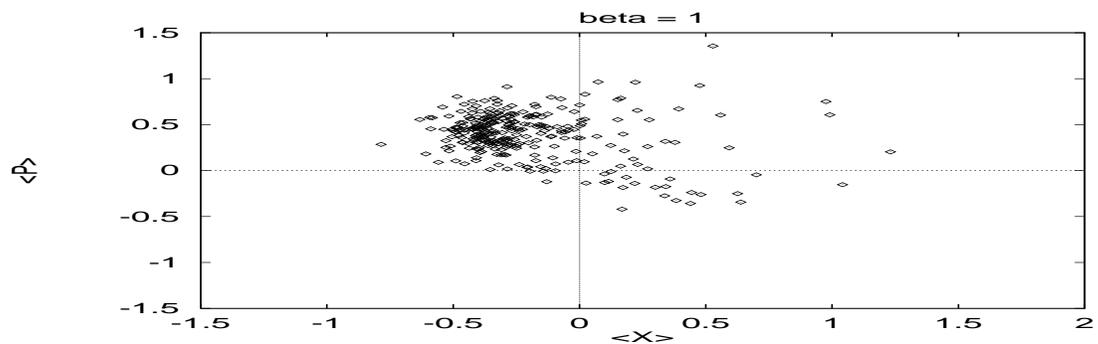
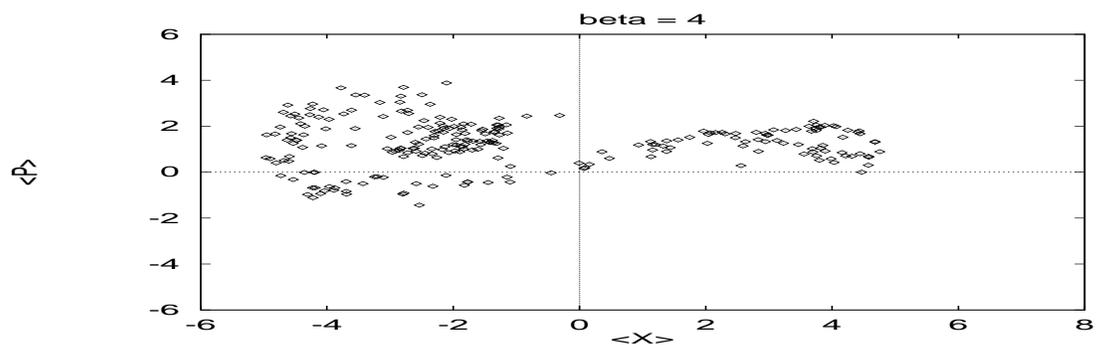
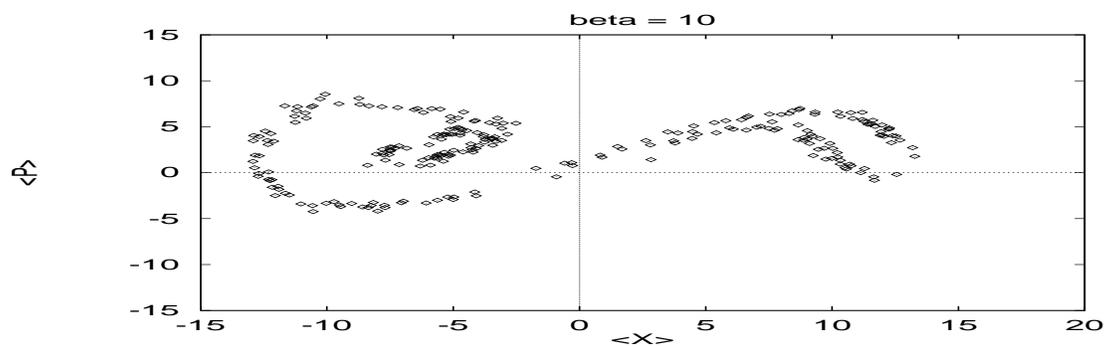
\end{document}